\documentclass[twocolumn,showpacs]{revtex4}

\begin{document}

%\preprint{}

\title{Gravitational contribution to fermion masses}

% Explanatory text should go in the []'s, actual e-mail

\author{A. Tiemblo}
\email[]{laetr03@imaff.cfmac.csic.es}
%\thanks{}
%\affiliation{}

\author{R. Tresguerres}
\email[]{romualdo@imaff.cfmac.csic.es}
%\thanks{}
\affiliation{Instituto de Matem\'aticas y F\'isica Fundamental\\
Consejo Superior de Investigaciones Cient\'ificas\\ Serrano 113
bis, 28006 Madrid, SPAIN}

\date{\today}

\begin{abstract}
In the context of a nonlinear gauge theory of the Poincar\'e
group, we show that covariant derivatives of Dirac fields include
a coupling to the translational connections, manifesting itself in
the matter action as a universal background mass contribution to
fermions.
\end{abstract}

%\keywords{}
\maketitle

\section{Introduction}
Conceived as an alternative to the standard general relativistic
metric approach to gravity, gauge theories of spacetime groups
describe gravitational forces in close analogy to the remaining
interactions \cite{Utiyama:1956sy} \cite{Sciama1} \cite{Sciama2}
\cite{Kibble:1961ba} \cite{Hayashi:1980wj} \cite{Lord:1987uq}
\cite{Lord:1988nd} \cite{Hehl:1995ue} \cite{Gronwald:1995em}. The
Lorentz group and the $GL(4\,,R)$ group are usual candidates
proposed by different authors \cite{Sciama1} \cite{Sciama2}
\cite{Ivanenko:1983vf} \cite{Sardanashvily:2002mi} to play the
role of local symmetries. Instead, Hehl et al.
\cite{Kibble:1961ba} \cite{Hehl:1995ue} \cite{Hehl:1974cn}
\cite{Hehl:1976kj} consider gravity as the gauge theory of either
the Poincaré\'e or the affine group: in any case of a group
including translations. Actually, the interpretation of tetrads as
a certain kind of translational connections allows an uniform
description of all known interactions, gravity included, in terms
of gauge potentials declared as the unique force mediators
\cite{Julve:1994bh} \cite{Lopez-Pinto:1995qb}
\cite{Lopez-Pinto:1997aw} \cite{Tresguerres:2000qn}.

We are interested in analyzing the consequences for matter fields
of considering translations included in the gauge group, as for
instance in the Poincar\'e gauge theory (PGT) of gravity, where
the full Poincar\'e group is treated as the local gauge group of a
Yang-Mills type theory. Given that such approach is a serious
candidate to become the fundamental theory of gravity, obviously
we must know how the corresponding Poincar\'e covariant
derivatives of matter fields look like, with both the homogeneous
Lorentz group contributions and those of translations taken into
account. The present paper is devoted to give an answer to the
question how translational connections couple in particular to
Dirac fields.

Independently of the interest of PGT in itself, the fact that we
choose it with preference to a more general gauge theory of
gravitation, such as metric-affine gravity (MAG), is partly
determined by a technical reason, namely the possibility of
building an explicit matrix representation of the Poincar\'e
algebra. In fact, besides the usual spin operators $\sigma
_{\alpha\beta}$ constituting the representation of the Lorentz
generators acting on Dirac fields, one can introduce the
complementary realization $\pi _\mu$ of the translational
generators. The affine group is more problematic to deal with due
to the fact that no finite dimensional spinor representation of
$GL(4\,,R)$ exists \cite{Hehl:1995ue}.

As an unexpected consequence of the explicit construction of
covariant Poincar\'e derivatives with intrinsic translations, we
find that the translational connections contribute to the Dirac
action with a fermion mass term of PGT-gravitational nature. Such
result is exclusive for a certain kind of gauge theories of
gravity, having nothing to do either with ordinary General
Relativity or with gauge approaches based on spacetime groups not
including translations. More precisely, we derive the background
fermion mass from the nonlinear approach to PGT established by us
in a number of previous papers \cite{Julve:1994bh}
\cite{Lopez-Pinto:1995qb} \cite{Lopez-Pinto:1997aw}
\cite{Tresguerres:2000qn} \cite{rrdph}. There we developed a
suitable treatment of spacetime groups with translations,
explaining the identification of tetrads as (nonlinear)
translational connections, and one of us proposed an adapted fibre
bundle description \cite{Tresguerres:2002uh}.

In next section we review a few main concepts, necessary to deduce
the key formula (\ref{compar4}) expressing nonlinear connections
in terms of linear ones. Then in section III we apply the
nonlinear procedure to the Poincar\'e group, paying special
attention to covariant derivatives of Dirac fields, and finally in
section IV we build the matter action showing the emergence of the
translation-induced mass term.

\section{Generalized bundle structure of gauge theories}

\subsection{Composite fiber bundles}
%\subsubsection{}

The ordinary gauge theory of a given Lie group $G$ is known to
have the structure of a principal bundle $P(M\,,G)$ equipped with
a connection, being matter fields defined on associated bundles.
However, gauge theories involving nonlinearly realized local
symmetries, as for instance gauge theories of spacetime groups,
require a slight modification of this bundle scheme, as discussed
in \cite{Tresguerres:2002uh}. The composite fiber bundles studied
there are particularly suitable to highlight the underlying
geometry of gauge theories of groups including translations, such
as the Poincar\'e gauge theory of gravity, thus constituting the
main support of the present paper. Let us briefly remaind the
reader on its defining features. For what follows, see
\cite{Sardanashvily:1992fq} \cite{Sardanashvily:1994}
\cite{Sardanashvily:1994fg} \cite{Sardanashvily:1995ew}, as much
as \cite{Kobayashi:1963}, pgs. 54 and 57.

Let $\pi _{_{PM}}:P\rightarrow M$ be a principal fiber bundle with
structure Lie group $G$, and let $H$ be a closed subgroup of $G$.
The quotient space $G/H$ constitutes a manifold on which $G$ acts
on the left in a natural way. Then it is possible to build the
$P$-associated bundle $\pi _{_{\Sigma M}}:\Sigma\rightarrow M$
with standard fiber $G/H$ and with total space consisting of the
quotient space $\Sigma =(P\times G/H)/G$ of the Cartesian product
$P\times G/H$ by the right action of $G$ defined as $P\times
G/H\ni (u\,,\xi\,)\rightarrow (ug\,,g^{-1}\xi\,)\in P\times
G/H\,,g\in G$. The total space $\Sigma$ can be identified with the
quotient space $P/H$ of $P$ by the right action of $H$ on $P$, and
consequently one finds $P(\Sigma\,,H)$ to be a principal fiber
bundle with structure group $H$ and with well defined projection
$\pi _{_{P\Sigma}}:P\rightarrow\Sigma$ onto the base space $\Sigma
=P/H$, see Prop. 5.5 of \cite{Kobayashi:1963}. Indeed, each orbit
$uH$ through $u\in P$ --diffeomorphic to the standard fiber $H$--
projects into a single element (a left coset) of $P/H$.

Nonlinearly realized gauge theories to be studied here differ from
ordinary gauge theories in that they are based on principal
bundles $P(M\,,G)$ whose structure group $G$ is reducible to a
closed subgroup $H$. According to Prop. 5.6 of
\cite{Kobayashi:1963}, such reducibility of the structure group
$G$ to $H\subset G$ is guaranteed if and only if a cross section
$s_{_{M\Sigma}}:M\rightarrow\Sigma =P/H$ of the associated bundle
$\Sigma$ exists. Furthermore, there is a one to one correspondence
between sections $s_{_{M\Sigma}}$ and the reduced subbundles of
$\pi _{_{P\Sigma}}:P\rightarrow\Sigma$ consisting of the set of
points $u\in P$ such that
\begin{equation}
\pi _{_{P\Sigma}}(u) =s_{_{M\Sigma}}\circ\pi
_{_{PM}}(u)\,,\label{subbundcondit}
\end{equation}
see \cite{Kobayashi:1963}. From condition (\ref{subbundcondit})
follows trivially
\begin{equation}
\pi _{_{PM}}=\pi _{_{\Sigma M}}\circ\pi
_{_{P\Sigma}}\,,\label{projdecomp}
\end{equation}
providing a decomposition of the total projection $\pi _{_{PM}}$
into partial projections. Accordingly, the principal bundle $\pi
_{_{PM}}:P\rightarrow M$ transforms into the composite bundle
\begin{equation}
\pi _{_{\Sigma M}}\circ\pi
_{_{P\Sigma}}:P\rightarrow\Sigma\rightarrow M\,.\label{compmanif}
\end{equation}
In (\ref{compmanif}) we distinguish two bundle sectors,
characterized respectively by the partial projections
\begin{equation}
\pi _{_{P\Sigma}}:P\rightarrow\Sigma\,,\quad\pi _{_{\Sigma
M}}:\Sigma\rightarrow M\,.\label{partproj}
\end{equation}
The latter one, with standard fiber $G/H$, can be seen as an
intermediate space, in the sense that it is built over the primary
base space $M$, and simultaneously plays a role as the base space
of the principal bundle $\pi _{_{P\Sigma}}:P\rightarrow\Sigma$
with structure group $H$. More precisely, in the context of
composite bundles one can regard the $G$-diffeomorphic fibers of
$P(M\,,G)$ as being, say, {\it bent} into two sectors,
corresponding respectively to the fibers $H$ of $\pi
_{_{P\Sigma}}:P\rightarrow\Sigma$ and $G/H$ of $\pi _{_{\Sigma
M}}:\Sigma\rightarrow M$. The $H$-diffeomorphic fiber branches are
attached to points of the {\it intermediate base space} $\Sigma$,
which trivialize locally as $(x\,,\xi )$, with $\xi$
coordinatizing the fiber branches $G/H$ attached to $x\in M$.

In parallel to (\ref{projdecomp}), the local sections
$s_{_{MP}}:M\rightarrow P$ are decomposed as
\begin{equation}
s _{_{MP}}=s _{_{\Sigma P}}\circ s
_{_{M\Sigma}}\,,\label{sectdecomp}
\end{equation}
see Section V of \cite{Tresguerres:2002uh}. In terms of suitable
zero sections, denoting as $\sigma _{_{MP}}$ those corresponding
to $s_{_{MP}}$, and so on, the sections in (\ref{sectdecomp})
become respectively
\begin{equation}
s_{_{MP}} =R_{\tilde{g}}\circ\sigma _{_{MP}}\,,\quad \tilde{g}\in
G \,,\label{locsect1}
\end{equation}
\begin{equation}
s_{_{\Sigma P}} =R_a\circ\sigma _{_{\Sigma P}}\,,\quad a\in H
\,,\label{locsect2}
\end{equation}
and
\begin{equation}
s_{_{M\Sigma}} =R_b\circ\sigma _{_{M\Sigma}}\,,\quad b\in G/H
\,.\label{locsect3}
\end{equation}
Conditions
\begin{equation}
\tilde{g} =b\cdot a\,,\quad \sigma _{_{\Sigma P}}\circ R_b
=R_b\circ\sigma _{_{\Sigma P}}\,,\label{condits}
\end{equation}
ensure that, in analogy to (\ref{sectdecomp}), the relation
\begin{equation}
\sigma _{_{MP}}=\sigma _{_{\Sigma P}}\circ\sigma
_{_{M\Sigma}}\label{zerosectdecomp}
\end{equation}
also holds. The usefulness of this structure will become evident
in the following.

In summary, composite fiber bundles (\ref{compmanif}) provide the
mathematical foundation for gauge theories involving nonlinear
gauge realizations (as the generalization of induced
representations). Relevant physical theories comprised among the
concerned ones are on the one hand the standard model --since a
correspondence between nonlinear realizations and spontaneous
symmetry breaking exists \cite{Cho:1978ss} -- and on the other
hand nonlinear gauge theories of gravity, as developed below.
Nonlinear realizations characteristic for such theories take place
on principal fiber bundles $P(M\,,G)$ whose structure group $G$ is
reducible to a closed subgroup $H\subset G$. While the total
symmetry remains that of the gauge group $G$, one exploits the
possibility of working with the explicit symmetry $H$ of the
principal subbundle of $\pi _{_{P\Sigma}}:P\rightarrow\Sigma$
whose base space is the total space of $\pi _{_{\Sigma
M}}:\Sigma\rightarrow M$. (The sections $s_{_{M\Sigma}}$ defined
on the latter bundle are identified as Goldstone fields
\cite{Sardanashvily:1992fq}.)

\subsection{Nonlinear realizations in composite bundles}

In \cite{Tresguerres:2002uh} it was showed that the composite
bundle structure defined by
(\ref{partproj})--(\ref{zerosectdecomp}) provides the natural
framework to deal with nonlinear gauge realizations, exactly as
standard principal bundles constitute the arena for the ordinary
gauge treatment of groups. Actually, the main results on nonlinear
realizations \cite{Coleman:1969sm} \cite{Callan:1969sn}
\cite{Salam:1969rq} \cite{Isham:1971dv} \cite{Borisov74}
\cite{Stelle:1980aj} are easily derived. So, the nonlinear gauge
transformation equation
\begin{equation}
L_g\circ\sigma _{_{\Sigma P}}(x\,,\xi )=R_h\circ\sigma _{_{\Sigma
P}}(x\,,\xi ')\label{nonlintrans1}
\end{equation}
is obtained by comparing two bundle elements, both with the form
(\ref{locsect2}), differing from each other by the left action
$L_g$ of elements $g\in G$, the latter being local in the sense
that $g=g(x)$, $x\in M$, see \cite{Tresguerres:2002uh} for
details. Regarding (\ref{nonlintrans1}) referred to the base space
$M$, it manifests itself as a vertical bundle automorphism not
affecting $x\in M$, in analogy to ordinary gauge transformations.
Nevertheless, when referred to the intermediate base space
$\Sigma\simeq M\times G/H$, the action of $L_g$ not only
transforms the sections $\sigma _{_{\Sigma P}}$ vertically along
the $H$ fiber branches by means of $R_h$, $h\in H$, but
simultaneously it induces a transformation affecting the points
$(x\,,\xi )\in\Sigma$, thus mapping $H$ fiber branches into fiber
branches defined on different $\Sigma$-points (as expected for
spacetime groups, in particular for translations, see
(\ref{varxi}) below).

In order to deal with ordinary geometrical objects defined on the
base manifold $M$, we pull back to the latter, by means of
$s_{_{M\Sigma}}$, the quantities defined on the plateau $\Sigma$.
Taking into account the property of pullbacks when applied to
functions $\varphi$, namely $f^*\varphi =\varphi\circ f$, we first
define the pullback of $\sigma _{_{\Sigma P}}$ as
\begin{equation}
\sigma _{\xi}(x):=(s_{_{M\Sigma}}^*\sigma _{_{\Sigma P}})(x)=
\sigma _{_{\Sigma P}}\circ s_{_{M\Sigma}}(x)\,.\label{sigmaxi}
\end{equation}
Then we calculate $s_{_{M\Sigma}}^* (L_g\circ\sigma _{_{\Sigma
P}})=L_g\circ\sigma _{_{\Sigma P}}\circ
s_{_{M\Sigma}}=L_g\circ\sigma _{\xi}$ and $s_{_{M\Sigma}}^*
(R_h\circ\sigma '_{_{\Sigma P}})=R_h\circ\sigma '_{_{\Sigma
P}}\circ s_{_{M\Sigma}}=R_h\circ\sigma _{\xi '}$, so that
(\ref{nonlintrans1}) gives rise to
\begin{equation}
L_g\circ\sigma _\xi (x)=R_h\circ\sigma _{\xi
'}(x)\,.\label{nonlintrans2}
\end{equation}
In (\ref{nonlintrans2}) (Eq.(6.6) of \cite{Tresguerres:2002uh})
one recognizes the fundamental equation for nonlinear realizations
\cite{Coleman:1969sm} \cite{Lord:1987uq} \cite{Lord:1988nd}.

The nonlinear gauge transformations of fields induced by
(\ref{nonlintrans2}) are deduced in Section VIII of
\cite{Tresguerres:2002uh}. Taking in (\ref{nonlintrans2})
$h\approx I+\mu$ to be infinitesimal, with $\mu$ defined on the
Lie algebra of $H$, the fields $\psi (\sigma _\xi (x)):=(\sigma
_\xi^*\,\psi )(x)$ of any given representation space of $H$ are
found to transform infinitesimally under $G$ as
\begin{equation}
\delta\psi (\sigma _\xi (x)):= \sigma _{\xi '}^*\,\psi  - (
L_g\circ\sigma _\xi )^*\,\psi\approx\rho (\mu)\psi (\sigma _\xi
(x))\,,\label{varmatt}
\end{equation}
being $\rho (\mu)$ the suitable representation of the $H$-algebra
element $\mu$. (See Eqs.(8.9), (8.11) of
\cite{Tresguerres:2002uh}.) Eqs. (\ref{nonlintrans2}) and
(\ref{varmatt}) summarize the main results of
\cite{Coleman:1969sm}. In the nonlinear approach, the relevant
fact is that the fields $\psi$ of representation spaces of
$H\subset G$ also constitute a representation space for the
nonlinear action (\ref{nonlintrans2}) of the full group $G$.

\subsection{Bundle approach to nonlinear connections and covariant
derivatives}

Covariant derivatives of the fields in (\ref{varmatt}) require the
introduction of suitable (nonlinear) connections on $M$. As a
crucial result for this purpose, in the present paragraph we will
derive equation (\ref{compar4}) below, implicit in
\cite{Tresguerres:2002uh} but not explicitly given there,
expressing the nonlinear connections in terms of standard (linear)
gauge potentials.

Depending on the bundle base space we consider, that is, either
$M$ or the {\it plateau} $\Sigma$, at least two alternative
expressions can be given for the Ehresmann connection form. On the
one hand, taking the quantities in (\ref{locsect1}) into account,
\begin{equation}
\omega =\tilde{g}^{-1}(\,d +\pi _{_{PM}}^*\,A_{_M} )\,\tilde{g}
\,,\label{connform1}
\end{equation}
involving the ordinary gauge potential $A_{_M}$ on the base space
$M$, defined as the pullback
\begin{equation}
A_{_M} =\sigma _{_{MP}}^*\,\omega\,.\label{linconn1}
\end{equation}
On the other hand, with (\ref{locsect2}) at view,
\begin{equation}
\omega =a^{-1}(\,d +\pi _{_{P\Sigma}}^*\,\Gamma _{_{\Sigma}}
)\,a\,,\label{connform2}
\end{equation}
where we introduce the nonlinear connection on the intermediate
space $\Sigma$, turning out to be the pullback
\begin{equation}
\Gamma _{_{\Sigma}}=\sigma _{_{\Sigma
P}}^*\,\omega\,.\label{nlinconn1}
\end{equation}
Since $\tilde{g} =b\cdot a$, see (\ref{condits}), comparison of
(\ref{connform1}) and (\ref{connform2}) yields
\begin{equation}
\pi _{_{P\Sigma}}^*\,\Gamma _{_{\Sigma}}= b^{-1}(\,d +\pi
_{_{PM}}^*\,A_{_M} ) \,b\,.\label{compar1}
\end{equation}
From the defining condition $\pi _{_{P\Sigma}}\circ\sigma
_{_{\Sigma P}}=id_{\Sigma}$ for sections follows $\sigma
_{_{\Sigma P}}^*\,\pi _{_{P\Sigma}}^*\,=id_{T^*\,(\Sigma )}$, so
that (\ref{compar1}) gives rise to
\begin{equation}
\Gamma _{_{\Sigma}}=\sigma _{_{\Sigma P}}^*\,[ \,b^{-1}(\,d +\pi
_{_{PM}}^*\,A_{_M} ) \,b\, ]\,.\label{compar2}
\end{equation}
We operate on (\ref{compar2}) taking into account that, in terms
of the pulled back quantity
\begin{equation}
\hat{b}(x\,,\xi ):=b\circ\sigma _{_{\Sigma P}}(x\,,\xi
)\,,\label{hatb}
\end{equation}
the relations $\sigma _{_{\Sigma P}}^*\,( b^{-1} d\,b)
=\hat{b}^{-1} d\,\hat{b}$, and $\sigma _{_{\Sigma P}}^*\,R_b^*\,=
R_{\hat{b}}^*\,\sigma _{_{\Sigma P}}^*$ hold, being $b^{-1}\pi
_{_{PM}}^*\,A_{_M}\, b =R_b^*\,\pi _{_{PM}}^*\,A_{_M}$ and, in
view of (\ref{projdecomp}), $\pi _{_{PM}}^*\,=\pi
_{_{P\Sigma}}^*\,\pi _{_{\Sigma M}}^*$. We find
\begin{equation}
\Gamma _{_{\Sigma}}=\hat{b}^{-1}(\,d +\pi _{_{\Sigma M}}^*\,A_{_M}
)\,\hat{b}\,.\label{compar3}
\end{equation}
Pulling back (\ref{compar3}), defined on $\Sigma$, by means of
$s_{_{M\Sigma}}^*$, compare with (\ref{sigmaxi}), we get
\begin{equation}
\Gamma _{_M}=s_{_{M\Sigma}}^*\,\Gamma
_{_{\Sigma}}\label{nlinconn2}
\end{equation}
as the nonlinear connection, defined on the base space $M$, we
will deal with in the following. Obviously, in view of
(\ref{nlinconn1}) and (\ref{sigmaxi})
\begin{equation}
\Gamma _{_M}=s_{_{M\Sigma}}^*\,\sigma _{_{\Sigma P}}^*\,\omega
=\sigma _\xi ^*\,\omega\,.\label{nlinconn3}
\end{equation}
Analogous calculations to those leading from (\ref{compar2}) to
(\ref{compar3}) allow us to find, in terms of the new pulled back
quantity
\begin{equation}
\tilde{b}(x):=b\circ\sigma _{_{\Sigma P}}\circ s_{_{M\Sigma}}(x)=b
(\sigma _\xi (x))\,,\label{tildeb}
\end{equation}
the relation
\begin{equation}
\Gamma _{_M}=\tilde{b}^{-1}( d + A_{_M}
)\,\tilde{b}\,,\label{compar4}
\end{equation}
between the nonlinear connection $\Gamma _M$ and the linear
connection $A_M$, that is, between the alternative pullbacks
(\ref{nlinconn3}) and (\ref{linconn1}) of the connection 1-form
$\omega$ to $M$. Our deduction of (\ref{compar4}) provides a
geometrical interpretation of eqs. (19) of \cite{Coleman:1969sm},
(6) of \cite{Cho:1978ss} and (2.7) of \cite{Tseytlin:1982nu},
while it shows the incompleteness of eqs. (2.15) of
\cite{Salam:1969rq} or (22) of \cite{Borisov74}. The importance of
(\ref{compar4}) for what follows becomes evident in view of the
transformation properties of $\Gamma _{_M}$, given in Eq.(7.14) of
\cite{Tresguerres:2002uh}, namely
\begin{equation}
\delta\Gamma _{_M}= \sigma _{\xi '}^*\,\omega - ( L_g\circ\sigma
_\xi )^*\,\omega\approx -(\,d\mu +[\Gamma _{_M}\,,\mu
])\,,\label{varnlcon}
\end{equation}
with $\mu$ being the same $H$-algebra element as in
(\ref{varmatt}). Eq. (\ref{varnlcon}) shows that only the
$H$-algebra components of $\Gamma _{_M}$ still transform
inhomogenously as $H$-connections, while the $G/H$-algebra
components transform as $H$-tensors. According to (\ref{varmatt})
and (\ref{varnlcon}), the nonlinear covariant differential defined
as
\begin{equation}
D\psi :=(d +\rho (\Gamma _{_M}))\psi\label{gencovder}
\end{equation}
transforms as an $H$-covariant differential
\begin{equation}
\delta D\psi =\rho (\mu )D\psi\label{vargencovder}
\end{equation}
under nonlinear gauge transformations (\ref{nonlintrans2}) of the
full group $G$. The general procedure established here will be
applied in the next section to $G$ as the Poincar\'e group and $H$
as the Lorentz group in order to derive PGT.

\section{Nonlinear Poincar\'e gauge theory of gravity}

\subsection{Poincar\'e covariant derivatives}

The main results of the previous section are summarized in the
transformation law (\ref{nonlintrans2}) and the induced field
transformation (\ref{varmatt}), plus the relation (\ref{compar4})
between the nonlinear connection (\ref{nlinconn3}) and the linear
one (\ref{linconn1}), with the corresponding nonlinear connection
transformation (\ref{varnlcon}). In terms of these elements, one
defines the covariant differential (\ref{gencovder}) transforming
as (\ref{vargencovder}).

Now, in order to perform explicit calculations, we need to
transform (\ref{nonlintrans2}) into a more manageable formula.
From (\ref{sigmaxi}) with (\ref{locsect3}), (\ref{condits}) and
(\ref{zerosectdecomp}), we get $\sigma _\xi (x) =R_b\circ\sigma
_{_{MP}}(x)$. (In the latter equation we identify $b=\sigma
_{_{MP}}^{-1}(x)\cdot\sigma _\xi (x)= b (\sigma _\xi
(x))=:\tilde{b}(x)$ as given by (\ref{tildeb}).) Analogously,
$\sigma _{\xi '}(x) =R_{b'}\circ\sigma _{_{MP}}(x)$. Replacing
these values into (\ref{nonlintrans2}), it follows $L_g\circ
R_b\circ\sigma _{_{MP}}(x)=R_h\circ R_{b'}\circ\sigma
_{_{MP}}(x)$. Finally, since $\sigma _{_{MP}}^{-1}(x)\cdot
g\cdot\sigma _{_{MP}}(x)=g$, we find
\begin{equation}
g\cdot b =b'\cdot h\,.\label{simplif}
\end{equation}
Eq. (\ref{simplif}) is the form of (\ref{nonlintrans2}) appearing
in \cite{Coleman:1969sm}, appropriate for practical computational
purposes, with $b$ being (\ref{tildeb}) and thus identical with
$\tilde{b}$ in (\ref{compar4}).

Now we merely apply the general formalism mechanically to the
gauge group $G=$ Poincar\'e, with $H=$ Lorentz. (Other choices of
$H$ have been studied elsewhere \cite{Lopez-Pinto:1997aw}.) In
(\ref{simplif}) we replace the infinitesimal group elements
$g\approx I+i\,\epsilon ^\mu P_\mu +i\,\beta ^{\alpha\beta}
L_{\alpha\beta}$ of the Poincar\'e group and $h\approx I+i\,\mu
^{\alpha\beta} L_{\alpha\beta}$ of the homogeneous Lorentz group,
and we parametrize $b$ and $b'$ respectively as $b=e^{-i\,\xi ^\mu
P_\mu}$ with finite translational parameters $\xi ^\mu$, and $b'
=e^{-i\,{\xi'}^\mu P_\mu}$ with ${\xi '}^\mu\approx\xi ^\mu
+\delta\xi ^\mu$. Then, taking into account the Poincar\'e
commutation relations
\begin{equation}
[\,L_{\alpha\beta}\,, L_{\mu\nu}]=-i\,( o_{\alpha [\mu}L_{\nu
]\beta} - o_{\beta [\mu}L_{\nu ]\alpha})\,,\label{comrel1}
\end{equation}
\begin{equation}
[\,L_{\alpha\beta}\,, P_\mu ]=i\,o_{\mu [\alpha}P_{\beta
]}\,,\quad [\,P_\mu\,, P_\nu ]=0 \,,\label{comrel2}
\end{equation}
with the help of the Hausdorff-Campbell formula, (\ref{simplif})
yields on the one hand the value $\mu ^{\alpha\beta} =\beta
^{\alpha\beta}$ for the $H$-parameter, and on the other hand
\begin{equation}
\delta\xi ^\mu =-\xi ^\nu \beta _\nu{}^\mu -\epsilon
^\mu\,.\label{varxi}
\end{equation}
Observe how the transformations (\ref{varxi}) of the translational
parameters resemble those of Cartesian coordinates.

Let us now pay attention to the connections. Starting with the
ordinary linear ones for the Poincar\'e group, say
\begin{equation}
A_{_M}=-i\,\Gamma ^{\alpha\beta}L_{\alpha\beta} -i\,{\buildrel
(T)\over{\Gamma ^\mu}}P_\mu\,,\label{linpoinc}
\end{equation}
we make use of (\ref{compar4}) to construct, in terms of
(\ref{linpoinc}) and of $b=e^{-i\,\xi ^\mu P_\mu}$, the nonlinear
connections
\begin{equation}
\Gamma _{_M}=-i\,\Gamma ^{\alpha\beta}L_{\alpha\beta}
-i\,\vartheta ^\mu P_\mu\,,\label{nlinpoinc}
\end{equation}
where simple calculations yield for the nonlinear translational
connection the structure
\begin{equation}
\vartheta ^\mu :=D\xi ^\mu + {\buildrel (T)\over{\Gamma
^\mu}}\,,\label{tetrad}
\end{equation}
being $D\xi ^\mu :=d\xi ^\mu +\Gamma _\nu{}^\mu \xi ^\nu$. More
explicitly, since all quantities are pulled back to the base space
$M$, (\ref{tetrad}) reads
\begin{equation}
\vartheta ^\mu =dx^i (\partial _i\xi ^\mu + \Gamma _{i\nu}{}^\mu
\xi ^\nu + {\buildrel (T)\over{\Gamma _i ^\mu}})=:dx^i
e_i{}^\mu\,,\label{tetradbis}
\end{equation}
where we introduce the usual notation $e_i{}^\mu$ for {\it
vierbeins} in order to show the identification we make of the
nonlinear translational connections with the tetrads. Such
interpretation of tetrads is possible since, in view of
(\ref{varnlcon}), they obey the gauge transformations
\begin{equation}
\delta\vartheta ^\mu =-\vartheta ^\nu \beta
_\nu{}^\mu\,.\label{vartetrad}
\end{equation}
In addition we find for the Lorentz part of (\ref{nlinpoinc})
\begin{equation}
\delta\Gamma _\alpha{}^\beta =D\beta
_\alpha{}^\beta\,.\label{varlorconn}
\end{equation}
As a consistence condition for (\ref{varxi}), (\ref{tetrad}),
(\ref{vartetrad}) and (\ref{varlorconn}) follows the
transformation of the linear translational connection
\begin{equation}
\delta {\buildrel (T)\over{\Gamma ^\mu}}=-{\buildrel
(T)\over{\Gamma ^\nu}}\beta _\nu{}^\mu +D\epsilon
^\mu\,.\label{vartransconn}
\end{equation}
Comparison of (\ref{vartransconn}) with the transformations
(\ref{vartetrad}) of the nonlinear translational connections
(\ref{tetrad}) clarify why the latter, as a result of the
nonlinear approach, can play the role of tetrads. Actually, tetrad
variations (\ref{vartetrad}) constitute a particular case of the
above mentioned fact that the nonlinear connection components
associated to generators of $G$ not belonging to $H$ behave as
$H$-tensors.

With the previous results at hand, the main task of the present
paragraph is to construct the Poincar\'e covariant derivatives of
matter fields. As shown by (\ref{varmatt}), the gauge action of
the full Poincar\'e group $G$ takes place through the
representation $\rho (\mu )=i\,\mu ^{\alpha\beta}\rho (
L_{\alpha\beta})$ of the algebra of the Lorentz group $H$, acting
on fields of arbitrary representation spaces of $H$. In
particular, for Dirac fields we take the spinor generators $\rho (
L_{\alpha\beta}) =\sigma _{\alpha\beta}$ as given by
(\ref{spinorgens}) below, being $\mu ^{\alpha\beta} =\beta
^{\alpha\beta}$ as mentioned just before (\ref{varxi}). We find
\begin{equation}
\delta\psi =i\beta ^{\alpha\beta}\sigma
_{\alpha\beta}\psi\,.\label{varpsi1}
\end{equation}
The covariant derivative (\ref{gencovder}) of such fields,
although resembling an ordinary $H$-covariant differential, is
build with a nonlinear connection defined on the whole
$G$-algebra. Thus, a representation of the full Poincar\'e algebra
is required in order to realize the nonlinear connection
(\ref{nlinpoinc}) as
\begin{equation}
\rho (\Gamma _{_M})=-i\,\Gamma ^{\alpha\beta}\sigma _{\alpha\beta}
-i\,\vartheta ^\mu \pi _\mu\,,\label{nlinpoincrep}
\end{equation}
where $\pi _\mu =\rho (P_\mu )$ is the finite matrix
representation of translational generators to be studied below.
According to the general formula (\ref{gencovder}), the Poincar\'e
covariant derivatives of Dirac fields read
\begin{equation}
D\psi =d\psi -i\,(\Gamma ^{\alpha\beta}\sigma _{\alpha\beta}
+\vartheta ^\mu \pi _\mu )\psi\,,\label{covder1}
\end{equation}
transforming in analogy to (\ref{varpsi1}) as
\begin{equation}
\delta D\psi =i\,\beta ^{\alpha\beta}\sigma
_{\alpha\beta}D\psi\,.\label{varcovder1}
\end{equation}
Certainly, due to the particular nonlinear Poincar\'e
transformations (\ref{vartetrad}) and (\ref{varpsi1}), the
contributions associated to the translational generators are not
necessary to guarantee covariance of (\ref{covder1}).
Nevertheless, the general scheme requires these contributions to
be present in the otherwise Lorentz covariant derivatives, as an
unavoidable heritage of the gauged Poincar\'e group. So we need to
know how the $G$ generators not belonging to $H$ act on the fields
$\psi$ of the representation space of $H$. In our case, this means
that, besides (\ref{spinorgens}), we have to look for the already
mentioned representation of the translational generators in order
to complete the finite matrix realization of the abstract
Poincar\'e algebra (\ref{comrel1}), (\ref{comrel2}).

\subsection{Intrinsic translations of fermion fields}

According to our conventions, the Dirac gamma matrices are defined
so that their product reads
\begin{equation}
\gamma _\alpha \gamma _\beta =-o_{\alpha\beta}\,I -4i\,\sigma
_{\alpha\beta}\,,\label{gammprod}
\end{equation}
expressed in terms of the Minkowski metric
\begin{equation}
o_{\alpha\beta}:=diag\,(-+++)\label{metric}
\end{equation}
and of the spinor generators
\begin{equation}
\sigma _{\alpha\beta}:={i\over8}\,[\,\gamma _\alpha\,,\gamma
_\beta ]\label{spinorgens}
\end{equation}
of the Lorentz group, being $\sigma _{\alpha\beta} =\rho (
L_{\alpha\beta})$ the usual $4\times 4$ matrix representation of
the Lorentz algebra (\ref{comrel1}) acting on 4-dimensional Dirac
bispinors $\psi$. Let us discuss how to extend the Lorentz algebra
to the Poincar\'e algebra, the latter one constituting a
subalgebra of the conformal algebra as shown in the appendix.

The possibility of constructing also intrinsic translational
operators $\pi _\mu =\rho ( P_\mu )$ from the gamma matrices rests
on the fact that
\begin{equation}
[\,\sigma _{\alpha\beta}\,,\gamma _\mu ]=i\,o_{\mu [\alpha}\gamma
_{\beta ]}\,,\label{commut2}
\end{equation}
and on the properties of the $\gamma _5$ matrix, defined as
\begin{equation}
\gamma _5:=i\,\gamma ^0\gamma ^1\gamma ^2\gamma
^3\,,\label{gamma5}
\end{equation}
such that $\gamma _5^2=I$ and satisfying the commutation relations
\begin{equation}
[\,\sigma _{\alpha\beta}\,,\gamma _5 ]=0\,,\label{commut3}
\end{equation}
and the anticommutation relations
\begin{equation}
\{\,\gamma _\mu\,,\gamma _5\}=0\,,\label{anticommut2}
\end{equation}
and
\begin{equation}
\{\,\sigma _{\alpha\beta}\,,\gamma _\mu \}=-{1\over2}\,\eta
_{\alpha\beta\mu}{}^\nu \gamma _\nu \gamma
_5\,,\label{anticommut1}
\end{equation}
(where $\eta _{\alpha\beta\gamma\delta}$, with $\alpha\,,\beta ...
=0,...,3$, is defined so that $\eta _{0abc}=\epsilon _{abc}$, with
$a,b,c=1,2,3$). Making use of these elements, one finds operators
\begin{equation}
\pi _\mu\sim \gamma _\mu (\,1+\lambda\,\gamma _5 )\label{linmom1}
\end{equation}
to exist, with $\lambda ^2=1$, satisfying the commutation
relations
\begin{equation}
[\,\sigma _{\alpha\beta}\,,\pi _\mu ]=i\,o_{\mu [\alpha}\pi
_{\beta ]}\,,\quad [\,\pi _\mu\,,\pi _\nu ]=0\,,\label{commut4}
\end{equation}
characteristic for translational generators, see (\ref{comrel2}).
Notice that eqs. (\ref{commut4}) do not completely determine $\pi
_\mu$. Actually, in (\ref{linmom1}) a global factor as much as the
sign $\lambda$ $(=\pm 1)$ remain unfixed. This fact reflects the
existence of two inequivalent realizations of the full conformal
algebra, of which the Poincar\'e algebra is a subalgebra. Invoking
dimensionality consistence of the intrinsic linear momentum $\pi
_\mu$ with the orbital linear momentum $-i\partial _\mu$ we
require the former, in natural units $\hbar =c=1$, to have
dimensions $[L]^{-1}$. Since the gamma matrices in (\ref{linmom1})
are dimensionless, we are enforced to introduce a dimensional
constant, say $m\sim [L]^{-1}$. Let us also fix the undetermined
sign in (\ref{linmom1}), see the appendix, and define
\begin{equation}
\pi _\mu :={m\over4}\,\gamma _\mu (\,1+\gamma _5
)\,,\label{linmom2}
\end{equation}
where the numerical factor is introduced for later convenience.

A remarkable feature of (\ref{linmom2}) is that $\pi _\mu \pi _\nu
=0$. The resulting anticommutation relations $\{\pi _\mu\,,\pi
_\nu\}=0$ are compatible with the finite matrix realization of the
Poincar\'e algebra given by (\ref{spinorgens}) and
(\ref{linmom2}). Since the commutation relations alone are
responsible for the transformations (\ref{varxi}) of the
coordinate-like parameters the matter fields depend on, they
suffice to induce the change from $\psi (\sigma _\xi (x))$ into
$\psi (\sigma _{\xi '} (x))$ where their gauge variation
(\ref{varmatt}) is evaluated.

On the other hand, the usual Casimir characterization of mass
still holds in our scheme despite the nilpotence of $\pi _\mu$ by
considering the complete translational generators as consisting of
the sum of an orbital plus an intrinsic contribution, namely
$P_\mu =iI\partial _\mu +\pi _\mu$. Observe that, in the limit of
vanishing components of the Lorentz connections, the translational
parameters $\xi ^\mu$ become indistinguishable from Cartesian
coordinates and the covariant derivative (\ref{covder1}) reduces
to the action of such a $P_\mu$ on fermions as $-i\,d\xi ^\mu
P_\mu\psi$. Since $\pi _\mu$ is traceless and $\pi _\mu\pi _\nu
=0$, the Casimir relation $Tr (P_\mu P^\mu)\sim m^2$ is valid for
$m\neq 0$.

Our intrinsic translational generators (\ref{linmom2}) resemble
the {\it momentum spin} introduced by G\"ursey \cite{Gursey:1964}
in the context of the contraction of $O(3\,,2)$ to the Poincar\'e
group \cite{Inonu:1953sp}. Indeed, such {\it momentum spin} is
conceived as the intrinsic part of the pseudotranslational
generators $\Pi _\mu := (1/R)\, L_{5\mu}$ whose commutation
relations, in the limit $R\rightarrow\infty$, reproduce those of
Poincar\'e translations.

\section{Poincar\'e gauge invariant Dirac action}

The discussion of previous section guarantees the translational
contributions in (\ref{covder1}) not only to make sense, but to be
an essential part of (nonlinear) Poincar\'e covariant derivatives.
Thus we have all the elements needed to build the Dirac matter
action in the presence of gravity, when the latter is described by
(nonlinear) PGT. Following the notation of \cite{Hehl:1990yq},
with $\gamma :=\vartheta ^\mu \gamma _\mu$, and $^*\gamma$ its
Hodge dual, the Dirac Lagrange density 4-form --without explicit
mass term-- reads
\begin{equation}
L_D={i\over2}\,\overline{\psi}\,\,^*\gamma\wedge D\psi +
h.c.\,,\label{lagrang1}
\end{equation}
with the usual definition $\overline{\psi}:=\psi ^\dagger\gamma
^0$, and $h.c.$ standing for the Hermitian conjugate of the given
term. Let us calculate the latter in order to make all our
conventions explicit. From (\ref{gammprod}) we get $\gamma
_0^2=1$. Provided
\begin{equation}
\gamma ^0\gamma ^\dagger _\mu\gamma ^0 =\gamma _\mu
\,,\label{gammadagg}
\end{equation}
as it is the case for instance for the Dirac representation of
gamma matrices in terms of Pauli matrices as
\begin{equation}
\gamma ^0 =\left(\begin{array}{ll}
I&\,\,0\\
0&-I
\end{array}\right)\,,\,
\gamma ^a =\left(\begin{array}{ll}
\,\,0&\sigma ^a\\
-\sigma ^a&0
\end{array}\right)\,,\,
\gamma _5 =\left(\begin{array}{ll}
0&I\\
I&0
\end{array}\right)\,,\label{matrices}
\end{equation}
we realize that
\begin{equation}
({i\over2}\,\overline{\psi}\,\,^*\gamma\wedge D\psi )^\dagger =
{i\over2}\,\overline{D\psi}\wedge\,^*\gamma\psi\,,\label{hermit}
\end{equation}
with $\overline{D\psi}:=(D\psi )^\dagger\gamma ^0$. Furthermore,
(\ref{gammprod}) with (\ref{gammadagg}) yields
\begin{equation}
\gamma ^0\sigma ^\dagger _{\alpha\beta}\gamma ^0 =\sigma
_{\alpha\beta}\,,\label{sigmadagg}
\end{equation}
guaranteeing the invariance of (\ref{lagrang1}) by enforcing
$\overline{\psi}$ to transforms as
\begin{equation} \delta\overline{\psi}=(\delta\psi )^\dagger\gamma
^0 =-i\overline{\psi}\,\beta ^{\alpha\beta}\sigma
_{\alpha\beta}\,,\label{varpsi2}
\end{equation}
and on the other hand from definition (\ref{gamma5}) with
(\ref{gammadagg}) we get
\begin{equation}
\gamma ^0\gamma ^\dagger _5\gamma ^0 =-\gamma
_5\,.\label{gamma5dagg}
\end{equation}
Applying (\ref{gammadagg}) and (\ref{gamma5dagg}) to
(\ref{linmom2}), it follows
\begin{equation}
\gamma ^0\pi ^\dagger _\mu\gamma ^0 =\pi _\mu\,,\label{pidagg}
\end{equation}
a result which was not {\it a priori} obvious. Taking
(\ref{sigmadagg}) and (\ref{pidagg}) into account, from
(\ref{covder1}) we find
\begin{equation}
\overline{D\psi}:=(D\psi )^\dagger\gamma ^0 = d\overline{\psi}
+i\,\overline{\psi}(\Gamma ^{\alpha\beta}\sigma _{\alpha\beta}
+\vartheta ^\mu \pi _\mu )\,,\label{covder2}
\end{equation}
transforming as
\begin{equation}
\delta\overline{D\psi}=-i\,\overline{D\psi}\,\beta
^{\alpha\beta}\sigma _{\alpha\beta}\,,\label{varcovder2}
\end{equation}
compare with (\ref{varpsi2}). If desired, in order to take into
account other forces besides gravitation, one can extend the gauge
symmetry replacing the Poincar\'e group by the direct product of
Poincar\'e times an internal group. To do so, one merely has to
replace (\ref{covder1}) by
\begin{equation}
D\psi =d\psi +i\,(gA -\Gamma ^{\alpha\beta}\sigma _{\alpha\beta}
-\vartheta ^\mu \pi _\mu )\psi\label{covder3}
\end{equation}
(and analogously (\ref{covder2})) without affecting what follows.
In view of the previous results, the explicit form of
(\ref{lagrang1}) becomes
\begin{equation}
L_D={i\over2}\,(\overline{\psi}\,\,^*\gamma\wedge D\psi
+\overline{D\psi}\wedge\,^*\gamma\psi )\,.\label{lagrang1bis}
\end{equation}
Let us separate the translational parts, no more indispensable for
the covariance of the covariant derivatives, from (\ref{covder1}),
(resp. (\ref{covder3})) as
\begin{equation}
D\psi =:\widetilde{D}\psi -i\vartheta ^\mu\pi
_\mu\psi\,,\label{covderdec1}
\end{equation}
and analogously
\begin{equation}
\overline{D\psi}=:\overline{\widetilde{D}\psi}
+i\,\overline{\psi}\,\vartheta ^\mu\pi _\mu\,,\label{covderdec2}
\end{equation}
see (\ref{covder2}), where we denote with tildes the
translations-independent pieces. Replacing (\ref{covderdec1}) and
(\ref{covderdec2}) in (\ref{lagrang1bis}), the Lagrange density
transforms into
\begin{equation}
L_D={i\over2}\,(\overline{\psi}\,\,^*\gamma\wedge\widetilde{D}\psi
+\overline{\widetilde{D}\psi}\wedge\,^*\gamma\psi )
+\,^*m\,\overline{\psi}\psi\,,\label{lagrang2}
\end{equation}
where we made use of the fact that $\vartheta
^\alpha\wedge\,^*\vartheta _\beta =\delta ^\alpha _\beta\,\eta $,
with $\eta =\,^*1$ as the 4-dimensional volume element, so that
\begin{equation}
^*\gamma\wedge\vartheta ^\mu \pi _\mu =-\eta\,\gamma ^\mu \pi _\mu
=\,^*m\,(1+\gamma _5 )\,,\label{calcul1}
\end{equation}
and
\begin{equation}
-\vartheta ^\mu \pi _\mu\wedge\,^*\gamma =-\eta\,\pi _\mu \gamma
^\mu =\,^*m\,(1-\gamma _5 )\,.\label{calcul2}
\end{equation}
Although $\gamma _5$ is necessary to guarantee the commutation
relations (\ref{commut4}) to hold, both contributions
(\ref{calcul1}) and (\ref{calcul2}) are combined in the action in
such a way that $\gamma _5$ cancels out. So the matter Lagrange
density (\ref{lagrang2}) merely retains a mass term, which is
unavoidable since it derives from the translational contribution
to the Poincar\'e connection (\ref{nlinpoincrep}). Accordingly,
either one of the projections $\psi _L$ or $\psi _R$ is lacking
(in which case $\overline{\psi}\psi =0$), or otherwise the field
$\psi$ is necessarily massive.

\section{Conclusions}

Independently from other possible origins of fermion masses, a
gravitational background mass contribution is predicted by PGT
when treated as a nonlinear local realization of the Poincar\'e
group. Provided both left and right projections of Dirac fields
are simultaneously present, (\ref{lagrang2}) prevents massless
Dirac fields from existing. The irremovable fermion masses are a
consequence of gravitational interaction (in particular of the
underlying translational group) in the context of PGT as the
fundamental theory of gravity.

As a phenomenological consequence, when considered together with
the standard model, PGT gives rise to a background contribution of
gravitational origin to the masses of all fermions: in particular
to the quark mass parameters of the QCD sector of the Lagrangian,
as much as to the neutrino masses. Neutrinos are thus predicted by
PGT to be massive. Certainly, our approach does not determine the
value of the universal translational mass parameter $m$. However,
from the observed masses of neutrinos it is clear that $m$ (the
same for all fermions) has to be very small, so that, accordingly,
its contribution to the observable hadron masses is expected to be
quite limited.

Matter currents corresponding to the Poincar\'e symmetry are the
spin current $\tau _{\alpha\beta}:={\partial L_D}/{\partial\Gamma
^{\alpha\beta}}$ and the energy-momentum 3-form $\Sigma _\mu
:={\partial L_D}/{\partial{\buildrel (T)\over{\Gamma ^\mu }}}
={\partial L_D}/{\partial\vartheta ^\mu }$. The former is found to
be $\tau _{\alpha\beta}=-{1\over4}\,\overline{\psi}\,\vartheta
_\alpha\wedge\vartheta _\beta\wedge\gamma\gamma _5\psi\,$ as it is
well known. Its coupling term to the Lorentz connection $\Gamma
^{\alpha\beta}$ falls off from the action in the limit of absence
of gravity (that is for $\Gamma ^{\alpha\beta} =0$, ${\buildrel
(T)\over{\Gamma ^\mu}}=0$). Instead, the mass term does not cancel
out in this limit. The reason is that, according to the nonlinear
approach to PGT, the tetrads have the structure (\ref{tetrad}),
not vanishing for zero linear connections. Actually, ordinary
Minkowskian flat spacetime may be regarded as the residual
structure left by nonlinear PGT in the absence of the
gravitational force carried by spin connections, that is in the
limit where the components of the latter ones are chosen to
vanish. The tetrads are in this case $\vartheta ^\mu =d\xi
^\mu\,$, so that the mass term associated to them still remains in
the action despite translational linear connections are switched
out.

% If in two-column mode, this environment will change to single-column
% format so that long equations can be displayed. Use
%\begin{widetext}
%\end{widetext}

\appendix*\section{The O(2\,,4) and the Poincar\'e algebra}

The Poincar\'e algebra is a subalgebra of the conformal algebra
\cite{Mack} to be examined here. Consider the $O(2\,,4)$
generators $L_{AB}=-L_{BA}$, $A\,,B...=0,...,3,5,6$, satisfying
the commutation relations
\begin{equation}
[\,L_{AB}\,, L_{MN}]=-i\,( g_{A[M} L_{N]B} - g_{B[M}
L_{N]A})\,,\label{o24commut}
\end{equation}
where the six-dimensional metric tensor is taken to be
\begin{equation}
g_{AB}=diag\,(-+++,+-)\,.\label{6dmetric}
\end{equation}
In order to relate (\ref{o24commut}) to the ordinary form of the
conformal commutation relations, let us decompose (\ref{6dmetric})
into the Minkowski metric
\begin{equation}
g_{\alpha\beta}=o_{\alpha\beta}:=diag\,(-+++)\,,\label{4dmetric}
\end{equation}
where $\alpha\,,\beta =0,...,3$, plus
\begin{equation}
g_{55}=1\,,\quad g_{66}=-1\,,\label{5,6dmetric}
\end{equation}
and define the translational generators
\begin{equation}
P_\mu :=L_{\mu 5}+ L_{\mu 6}\,,\label{translat}
\end{equation}
the special conformal generators
\begin{equation}
K_\mu :=L_{\mu 5}- L_{\mu 6}\,,\label{specconf}
\end{equation}
and the dilatational generators
\begin{equation}
D :=-2\,L_{56}\,.\label{dilat}
\end{equation}
In terms of $L_{\alpha\beta}$, (\ref{translat}), (\ref{specconf})
and (\ref{dilat}), the commutation relations (\ref{o24commut})
give rise to the conformal algebra
\begin{equation}
[\,L_{\alpha\beta}\,, L_{\mu\nu}]=-i\,( o_{\alpha [\mu} L_{\nu
]\beta} - o_{\beta [\mu} L_{\nu ]\alpha})\,,\label{confcommut0}
\end{equation}
\begin{equation}
[\, L_{\alpha\beta}\,,P_\mu ]=i\,o_{\mu [\alpha}P_{\beta
]}\,,\label{confcommut1}
\end{equation}
\begin{equation}
[\, L_{\alpha\beta}\,,K_\mu ]=i\,o_{\mu [\alpha}K_{\beta
]}\,,\label{confcommut2}
\end{equation}
\begin{equation}
[\,P_\mu\,,K_\nu ]=i\,( L_{\mu\nu}
+{1\over2}\,o_{\mu\nu}D)\,,\label{confcommut3}
\end{equation}
\begin{equation}
[\,D\,,P_\mu ]=-i\,P_\mu\,,\label{confcommut4}
\end{equation}
\begin{equation}
[\,D\,,K_\mu ]=i\,K_\mu\,,\label{confcommut5}
\end{equation}
\begin{equation}
[\,P_\mu\,,P_\nu ]=[\,K_\mu\,,K_\nu ]=[\,D\,,
L_{\mu\nu}]=[\,D\,,D]=0\,.\label{confcommut6}
\end{equation}
As pointed out in \cite{Mack}, all finite dimensional
representations of the $O(2\,,4)$ algebra can be obtained by
reducing out tensor products of two inequivalent fundamental
4-dimensional representations (corresponding respectively to the
choices $\lambda =\pm 1$ in what follows) builded from the gamma
matrices as
\begin{equation}
\rho ( L_{\alpha\beta})=\sigma _{\alpha\beta}
:={i\over8}\,[\,\gamma _\alpha\,,\gamma _\beta ]\,,\label{Lorgens}
\end{equation}
\begin{equation}
\rho ( L_{\mu 5})={1\over2}\,(\pi _\mu +\kappa _\mu ) =\lambda
{m\over4}\,\gamma _\mu\gamma _5\,,\label{mu5gens}
\end{equation}
\begin{equation}
\rho ( L_{\mu 6})={1\over2}\,(\pi _\mu -\kappa _\mu ) =
{m\over4}\,\gamma _\mu\,,\label{mu6gens}
\end{equation}
\begin{equation}
\rho ( L_{56})=-{1\over2}\,\Delta =-\lambda\,{i\over4}\,\gamma
_5\,.\label{56gens}
\end{equation}
Obviously, as read out from (\ref{mu5gens}), (\ref{mu6gens}) and
(\ref{56gens}), the corresponding fundamental inequivalent
representations of (\ref{translat}), (\ref{specconf}) and
(\ref{dilat}) read
\begin{equation}
\pi _\mu :={m\over4}\,\gamma _\mu (\,1+\lambda\gamma _5
)\,,\label{transgens}
\end{equation}
\begin{equation}
\kappa _\mu :=-{m\over4}\,\gamma _\mu (\,1-\lambda\gamma _5
)\,,\label{specconfgens}
\end{equation}
and
\begin{equation}
\Delta :=\lambda \,{i\over2}\,\gamma _5\,,\label{dilgens}
\end{equation}
where the role of $\pi _\mu$ and $\kappa _\mu$ is interchangeable
by fixing $\lambda$ to be either $\pm 1$, and by accordingly
change the sign of (\ref{dilgens}). The Poincar\'e algebra
considered in the main text is the subalgebra of the conformal
algebra consisting of the spin and translational generators only,
having fixed $\lambda =1$.

\begin{acknowledgments}
The authors are very grateful to Friedrich Wilhelm Hehl and to
Yuri Obukhov for helpful and clarifying discussions.
\end{acknowledgments}


\end{document}